\documentclass[twocolumn, showpacs,preprintnumbers,amsmath,amssymb]{revtex4}
\usepackage{graphicx}
\usepackage{dcolumn}
\usepackage{bm}

\begin{document}

\title{An effective packing fraction for better resolution near the critical point of shear thickening suspensions}

\author{Rijan Maharjan} 
\author{Eric Brown}

\affiliation{Department of Mechanical Engineering and Materials Science, Yale University, New Haven, CT 06520}
\date{\today} 

\begin{abstract}
We present a technique for obtaining an effective packing fraction for discontinuous shear thickening suspensions near a critical point.  It uses a measurable quantity that diverges at the critical point -- in this case the inverse of the shear rate $\dot\gamma_c^{-1}$ at the onset of discontinuous shear thickening --  as a proxy for packing fraction $\phi$.  We obtain an effective packing fraction for cornstarch and water by fitting $\dot\gamma_c^{-1}(\phi)$, then invert the function to obtain $\phi_{eff}(\dot\gamma_c)$.  We further include the dependence of $\dot\gamma_c^{-1}$ on the rheometer gap $d$ to obtain the function $\phi_{eff}(\dot\gamma_c,d)$.  This effective packing fraction $\phi_{eff}$ has better resolution near the critical point than the raw measured packing fraction $\phi$ by as much as an order of magnitude.  Furthermore, $\phi_{eff}$ normalized by the critical packing fraction $\phi_c$ can be used to compare rheology data for cornstarch and water suspensions from different lab environments with different temperature and humidity.  This technique can be straightforwardly generalized to improve resolution in any system with a diverging quantity near a critical point.
\end{abstract}

\maketitle

\section{Introduction}
\label{sec:Intro}

Many densely packed suspensions such as cornstarch and water are known to exhibit Discontinuous Shear Thickening (DST).  DST is defined by an effective viscosity function $\eta$ that increases apparently discontinuously as a function of shear rate $\dot\gamma$ (for reviews, see \cite{Ba89, WB09, BJ14, DMB18}).    DST fluids exhibit a number of unusual phenomena, such as the ability to support a person running or walking on the surface \cite{MAB18}, giant fluctuations in stress \cite{LDH03, LDHH05}, hysteretic flow and oscillations \cite{De10, KSLM11, KSM13, NNM12, PCWB15}, shear-induced jamming \cite{WJ12, WRVJ13, PJ14, HPJ16}, and anomalous relaxation times \cite{MB17}.  These phenomena and DST tend to be found at packing fractions $\phi$ just a few percent below the packing fraction of the liquid-solid transition $\phi_c$ (a.k.a.~jamming).  $\phi_c$ is a critical point,  in the sense that the magnitude of the viscosity and steepness of the shear thickening portion of the $\eta(\dot\gamma)$ curve in stress-controlled measurements diverge in the limit as the packing fraction $\phi$ is increased to $\phi_c$ \cite{KD59, BJ09, BZFMBDJ11}.  Near such a critical point, any uncertainty on the control parameter (in this case packing fraction $\phi$) can lead to enormous uncertainties in output parameters that are sensitive to the control parameter (e.g.~viscosity magnitude or slope \cite{BJ09, BZFMBDJ11}, jamming front propagation speeds \cite{WRVJ13}, and relaxation times \cite{MB17}), making it challenging to study trends in packing fraction in this range and identify how these phenomena are related to each other or to DST. 

Measurements of packing fractions of suspensions typically have uncertainties around 0.01 \cite{BJ09,BZFMBDJ11}, corresponding to $20\%$ of the DST range \cite{MB17}.  For cornstarch, this error comes partly from adsorption of water from the air onto the particles.   While a suspension is being mixed, placed, and measured it adsorbs water from the air, and water evaporates, depending on temperature and humidity.   For example, a variation of 5\% in relative humidity results in an error of 0.01 in $\phi$ for cornstarch and water at equilibrium \cite{SF44}.  Even samples that don't interact with water in the atmosphere typically have random uncertainties in packing fraction around 0.01 \cite{BJ09, BZFMBDJ11}.  This can result from the difficulty of loading a sample onto a rheometer from a mixer without changing the proportion of particles to solvent.  
  

The sensitivity of $\phi$ to temperature and humidity  can also result in a huge systematic error when comparing to data from different labs, or from different seasons in the same lab where the humidity is different.   For example, a container of nominally dry cornstarch can be between 1\% and 20\% water at equilibrium depending on humidity and temperature where it is stored \cite{SF44}.  This, and along with different packing fraction measurement techniques, can result in systematic differences between different labs in reported values of $\phi_c$ and thus $\phi$ of around 0.1 \cite{BJ09, FHBOB08, MB17} --   larger than the 0.05 range in $\phi$ where DST is found.   Without a common packing fraction scale, datasets for cornstarch suspensions from different labs have remained for the most part uncomparable.  

The random and systematic errors can be reduced by using as a reference a measurable quantity that diverges at the critical point, which can be converted to an effective packing fraction $\phi_{eff}$.  In a previous work, we used the inverse of the shear rate at the onset of DST  $\dot\gamma_c^{-1}$ as a reference that allowed measurement of trends in a relaxation time in the range $\phi_c-0.02 < \phi_{eff} < \phi_c$ for cornstarch and water \cite{MB17}.  In this methods paper, we expand on the technique we introduced previously \cite{MB17} to include the dependence of the effective packing fraction on both onset shear rate $\dot\gamma_c$ and rheometer gap $d$, since $\dot\gamma_c$ is known to depend $d$ \cite{FHBOB08, FBOB12}.  This produces a more generally useful conversion function which any experimenter in any laboratory conditions could use to obtain a value of $\phi_{eff}$ for cornstarch and water, requiring only measurements of the critical shear rate $\dot\gamma_c$ and the gap $d$.   We also show how much the precision on packing fraction measurements is improved using this technique.  While we present this technique using a shear thickening transition as an example, the technique could in principle be applied to improve resolution for any system with measurable quantities that diverge at a critical point.

The remainder of this manuscript is organized as follows.  The materials and experimental methods used are given in Secs.~\ref{sec:materials} and \ref{sec:methods}, respectively.  Examples of viscosity curves are shown in Sec.~\ref{sec:viscosity}.  The method to calculate the critical shear rate $\dot\gamma_c$ from the viscosity curves in shown in Sec.~\ref{sec:dotgammac}.  Fits to 
 obtain $\phi_{eff}(\dot\gamma_c)$  are shown in Sec.~\ref{sec:phieff_dotgammac}, and fits of $\dot\gamma_c(d)$ are shown in Sec.~\ref{sec:gapdependence}, which are combined to obtain $\phi_{eff}(\dot\gamma_c, d)$ in Sec.~\ref{sec:combined_model}.  Section \ref{sec:scatter} compares the random errors on $\phi_{eff}$ to those on $\phi_{wt}$, which identifies the range of packing fraction where $\phi_{eff}$ has an improved error over $\phi_{wt}$.  Section \ref{sec:systematicerror} shows how systematic errors are improved by using $\phi_{eff}/\phi_c$.

\section{Materials}
\label{sec:materials}

The suspensions used were the same as our previous work \cite{MB17}.  Cornstarch was purchased from \textit{Carolina Biological Supply} and suspended in tap water,  to obtain a typical DST fluid \cite{BJ14}. The samples were created at a temperature of $22.0\pm0.6$ $^{\circ}$C and humidity of $48\pm6\%$, where the uncertainties represent day-to-day variations in the respective values.  A four-point scale was used to measure quantities of cornstarch and water to obtain a weight fraction $\phi_{wt}$. 

Each suspension was stirred until no dry powder was observed. The sample was further shaken in a \textit{Scientific Instruments Vortex Genie 2} for 30 seconds to 1 minute on approximately 60\% of its maximum power output. 

\section{Experimental methods}
\label{sec:methods}

The experimental methods used are identical to our previous work \cite{MB17}. Suspensions were measured in an \textit{Anton Paar MCR 302} rheometer in a parallel plate setup. The rheometer measured the torque $M$ on the top plate  and  angular rotation rate $\omega$  of the top plate.   The mean shear stress is given by $\tau = 2M/\pi R^3$ where $R$ is the radius of the sample.  While the mean shear rate varies along the radius of the suspension, the mean shear rate at the edge of the plate is used as a reference parameter, which is given by $\dot{\gamma} = R\omega/d$ where $d$ is the size of the gap between the rheometer plates, equal to the sample thickness.  The viscosity of the sample is measured as $\eta = \tau/\dot{\gamma}$ in a steady state.  We took two data series with approximately fixed gaps $d=1.250$ mm and $d=0.610$ mm.  Within a series, we allowed $d$ to vary with a standard deviation of up to  0.024 mm from experiment to experiment in an attempt to reduce the uncertainty on the sample radius $R=25.0\pm0.5$ mm.   The experiments were performed at a plate temperature of $23.5\pm0.5$ $^{\circ}$C. A solvent trap was used to slow down the moisture exchange between the sample and the atmosphere. The solvent trap effectively placed a water seal around the  sample, with a lipped lid around the sample and the lips touching a small amount of water contained on the top, cupped, surface of the tool.

We pre-sheared the sample before viscosity curve measurements to reduce effects of loading history. The preshear was performed over 200 s with a linear ramp in shear rate, covering the entire range of shear rates with shear thickening for higher weight fractions (what we later identify to be $\phi_{eff} \ge 0.547$), and the measurable range of shear thickening for lower $\phi_{eff}$ (this was limited by spillage of the sample at higher shear rates). The net strain on the sample was at least 10 over the course of the preshear, ensuring a well-developed sheared structure.  We measured viscosity curves immediately after this pre-shear by ramping the shear rate down then up to minimize acceleration of the sample, performing the ramps twice in sequence.  The shear rate was ramped at a rate of  250 to 500 seconds per decade of shear rate, which was slow enough to obtain reproducible viscosity curves without hysteresis within a typical run-to-run variation of 30\%.

\section{Obtaining the effective packing fraction $\phi_{eff}$}
\label{sec:results}

\subsection{Steady state viscosity curves}
\label{sec:viscosity}

\begin{figure}
\centering
\includegraphics[width=0.475\textwidth]{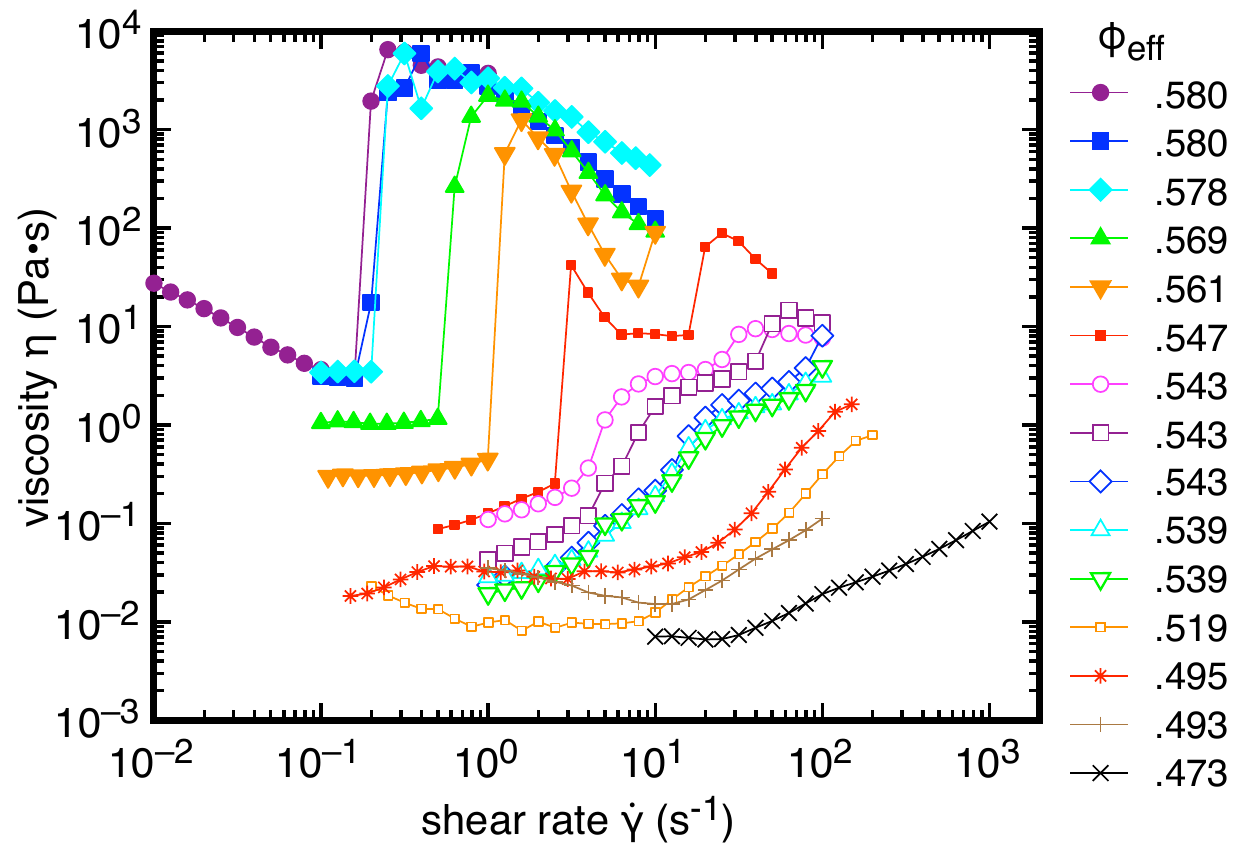}
\caption{(color online) Viscosity $\eta$ as a function of shear rate $\dot\gamma$, at  different effective packing fractions $\phi_{eff}$ shown in the key.   Solid symbols: discontinuous shear thickening (DST) range.  Non-solid symbols:  continuous shear thickening range. 
}
\label{fig:viscosity_shearrate_all}
\end{figure}

To obtain an effective packing fraction $\phi_{eff}(\dot\gamma_c)$, we need to obtain the onset shear rate $\dot\gamma_c$ from viscosity curves.  Figure \ref{fig:viscosity_shearrate_all} shows curves of viscosity $\eta$ as a function of shear rate $\dot\gamma$ for $d=0.610$ mm.  Each of these curves is an average of the four ramps measured.  Different packing fractions are represented by the values of $\phi_{eff}$ in the legend of Fig.~\ref{fig:viscosity_shearrate_all} (These are obtained from Eq.~\ref{eqn:combined_model} which will be explained in Sec.~\ref{sec:combined_model}).  Shear thickening is defined by the regions of positive slope of $\eta(\dot\gamma)$.  For $\phi_{eff} >0.547$ (solid symbols in Fig.~\ref{fig:viscosity_shearrate_all}), sharp jumps in $\eta$ are observed at a critical shear rate $\dot\gamma_c$.  Such sharp jumps are usually identified as discontinuous shear thickening (DST).  The shear thickening is relatively weak at lower $\phi_{eff}$ (i.e.~the slope is shallower), which is usually identified as continuous shear thickening.  For $\phi_{wt} > 0.610\pm0.007 = \phi_c$, we observe a large yield stress even at very low shear rates (not shown here) \cite{MB17}, corresponding to a solid (a.k.a.~jammed) state. The examples here are similar to other examples of shear thickening in the literature \cite{Ba89, WB09, BJ14, DMB18}.

\subsection{Method to obtain $\dot\gamma_c$}
\label{sec:dotgammac}

We define $\dot\gamma_c$ at the onset of DST for the average of the four viscosity curves, where the viscosity increase is the sharpest near the onset of shear thickening.  Identifying this onset is trivial for discontinuous-looking curves, as the increase is very sharp.  At lower $\phi_{eff}$, there is no sharp transition, but a more gradual increase in the slope of $\eta(\dot\gamma)$.   To account for both of these regimes, we identify $\dot\gamma_c$ as the average of the smallest adjacent pair of $\dot\gamma$ values where the local slope $\partial \eta/\partial \dot\gamma > 1$.  For weaker shear thickening, this condition on the slope is never met, so $\phi_{eff}$ is not defined, although $\phi_{eff}$ would likely be less useful so far away from the critical point anyway.  
 We define the viscosity $\eta_c$ at the onset of DST as the viscosity at the lower of the two shear rates used for $\dot\gamma_c$ as representative of the viscosity on the lower side of the shear thickening  transition.  This method of averaging the four viscosity curves before finding $\dot\gamma_c$ leads to more consistent ordering of  viscosity curves shown in Fig.~\ref{fig:viscosity_shearrate_all} in terms of increasing $\eta_c$ and $\dot\gamma_c^{-1}$ with $\phi_{eff}$, compared to calculating $\dot\gamma_c$  individually for each viscosity ramp then averaging over multiple ramps as done in our previous work \cite{MB17}.  
 
 The run-to-run variation can be characterized by the standard deviation of the four ramps, which is on average 32\% for $\dot\gamma_c$ and $28\%$ for $\eta_c$ for $\phi_{eff}<0.54$, similar to the run to run variation in viscosity \footnote{The error on $\eta_{min}$ was erroneously reported to be larger in Ref.~\cite{MB17}, although the conclusions of that analysis would not change with the smaller errors, and the error analysis is more thoroughly presented in this paper}.  For smaller $\phi_{eff}$, the slopes of the viscosity curves become closer to the threshold $\partial \eta/\partial \dot\gamma = 1$, so noise in the data causes errors on the calculation of $\eta_c$ that tend to be larger than the typical run-to-run variation of viscosity of 30\%. We will show in Sec.~\ref{sec:scatter} that the effective packing fraction does not  improve resolution over $\phi_{wt}$ for $\phi_{eff}< 0.54$ anyway.


\subsection{$\phi_{eff}(\dot\gamma_c)$}
\label{sec:phieff_dotgammac}

\begin{figure}
\centering
\includegraphics[width=0.45\textwidth]{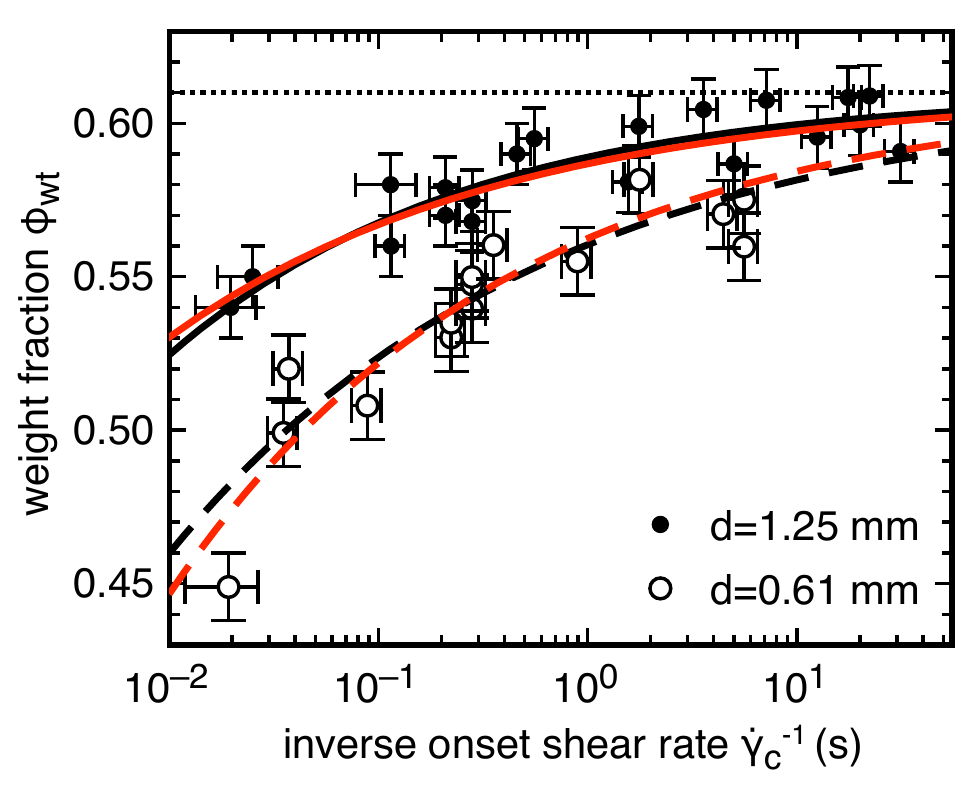}
\caption{(color online) Weight fraction $\phi_{wt}$ as a function of the inverse shear rate $\dot\gamma_c^{-1}$ at the onset of DST.   Solid symbols: gap $d = 1.25$ mm. Open symbols:  $d = 0.61$ mm. Black lines: power law fit for $d = 1.25$ mm (solid line) and for $d=0.61$ mm (dashed line).  Red lines: model curves for $\phi_{eff}$ corresponding to the best simultaneous fit of Eq.~\ref{eqn:combined_model} to the data at both gaps $d$.  Dotted line: critical packing fraction $\phi_c = 0.610$.
}
\label{fig:phi_cricSR_all}
\end{figure} 

To obtain the function for $\phi_{eff}(\dot\gamma_c)$ we fit of $\phi_{wt}(\dot\gamma_c$) \cite{MB17}.   Figure \ref{fig:phi_cricSR_all} shows a plot of $\phi_{wt}(\dot\gamma_c$) for two different gaps $d$.   The data for $d=1.25$ mm are reproduced from our previous work with the same experimenter and methods \cite{MB17}.   The data are plotted in terms of $\dot\gamma_c^{-1}$ so that the effective packing fraction increases from left to right.  The fact that the two sets of data do not collapse confirms that there is a dependence of $\dot\gamma_c$ on gap size \cite{FBOB12}.  To obtain a conversion function $\phi_{wt}(\dot\gamma)$, we least-squares fit a power law 
\begin{equation}
\phi_{wt} = A\dot\gamma_c^{B}+\phi_c
\label{eqn:phidependence}
\end{equation}

\noindent to the data with fit parameters $A$ and $B$.  The black lines in Fig.~\ref{fig:phi_cricSR_all} show least squares fits of Eq.~\ref{eqn:phidependence} to each set of data with a fixed $d$.    We fixed $\phi_c=0.610\pm0.007$ at the value of the jamming transition $\phi_c$ (where the yield stress is non-zero for $\phi>\phi_c$), as that was obtained from a best fit of the same function \cite{MB17}, and the same value of $\phi_c$ is expected for different $d$ as long as $d$ is more than a few particle diameters \cite{DW09, BZFMBDJ10}.  Since the onset stress of DST is mostly independent of packing fraction \cite{BJ14}, the divergence of viscosity with packing fraction leads to the divergence of $\dot\gamma_c^{-1}$ in the limit as $\phi_{wt}$ approaches $\phi_c$ \cite{BJ09, MB17}, and the exponent $B$ has the same meaning as the exponent in the Krieger-Doherty relation \cite{KD59}.  We use the standard deviation of the mean on $\dot{\gamma}_c$ of 16\% as an input error.  We also adjust errors in $\phi_{wt}$ to a constant value of 0.008 to obtain a reduced $\chi^2= 1$.  The input error of 0.008 indicates a combination of the sample-to-sample uncertainty on $\phi_{wt}$ for our measurements plus any deviation of the fit function from the `true' function describing the data.  The fit yields $A=-0.0210\pm0.0022$ (for $\dot\gamma_c^{-1}$ in units of seconds) and $B=0.303\pm 0.038$ for $d=1.25$ mm.  As a self-consistency check, if  we instead  additionally fit the value of $\phi_c$, then we obtain $\phi_c=0.609\pm 0.008$ without significant reduction in $\chi^2$.  This is consistent with the value of $\phi_c=0.610\pm0.007$ obtained from an earlier fit \cite{MB17}, as well measurements of a yield stress at $\phi > \phi_c$ \cite{MB17}.  The fit of Eq.~\ref{eqn:phidependence} to data at $d=0.61$ mm shown in Fig.~\ref{fig:phi_cricSR_all} yields $A=-0.0495\pm0.0038$ (for $\dot\gamma_c$ is in units of seconds) and $B=0.241\pm 0.031$, with a sample-to-sample uncertainty of 0.011 required to obtain a reduced $\chi^2=1$.  These exponents $B$ for the different gaps $d$ are consistent with each other within their errors, while the different coefficients $A$ are are a result of the $d$-dependence of $\dot\gamma_c$ \cite{FHBOB08, FBOB12}.   We define $\phi_{eff}(\dot\gamma_c)$ using the best fit parameters from Eq.~\ref{eqn:phidependence} with $\phi_{eff}$ in place of $\phi_{wt}$ so that our effective packing fraction is based on the measurement of $\dot\gamma_c$, but is still can be interpreted as a packing fraction with a value close to $\phi_{wt}$ \cite{MB17}.


\subsection{Gap dependence}
\label{sec:gapdependence}

\begin{figure}
\centering
\includegraphics[width=0.45\textwidth]{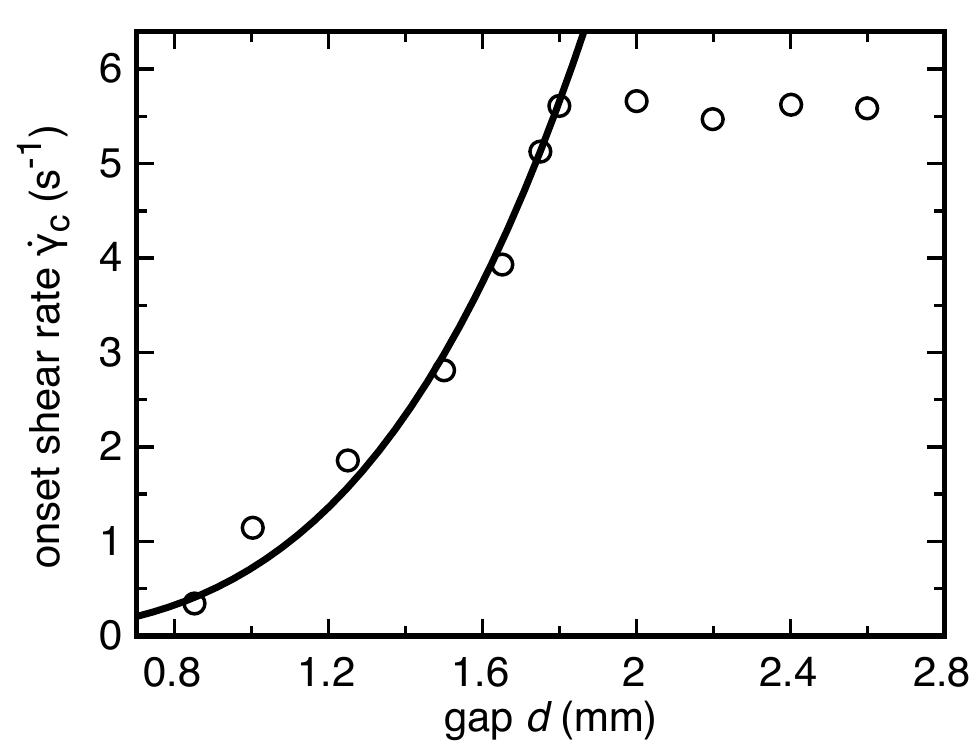}
\caption{Onset shear rate $\dot\gamma_c$ as a function of gap $d$.  Data is reproduced from Fall et al.~\cite{FBOB12}.  Solid line: best fit of Eq.~\ref{eqn:gapdependence} for $d \le 1.8$ mm.
}
\label{fig:bonn_check}
\end{figure}

We saw that the onset shear rate $\dot\gamma_c$ depends on the rheometer gap size $d$, as found previoysly \cite{FHBOB08, FBOB12}. In order to obtain a more complete conversion function $\phi_{eff}(\dot\gamma_c,d)$, we fit the data of Fall et al.~\cite{FBOB12}.  Fall et al.~\cite{FBOB12} presented the dependence of the onset shear rate $\dot\gamma_c$ on the gap $d$ for a parallel plate tool of radius $R=20$ mm.   They used a density matched suspension of cornstarch in a 55 wt\% solution of CsCl in demineralized water at a nominal packing fraction $\phi_{wt}=0.44$ (which is in the DST range on their scale).   While the use the CsCl for density matching is different than the solvent used in our measurements, to our knowledge there is no qualitative effect of density matching on shear thickening measurements of cornstarch and water.   The solvent viscosity is also somewhat larger with CsCl.  Since we only want to obtain the scaling of $\dot\gamma_c(d)$ from this data, we assume and confirm later that the scaling in $d$ is independent of these parameter values.  

The onset shear rate $\dot\gamma_c$ as a function of gap $d$ from Fall et al.~\cite{FBOB12} is reproduced in Fig.~\ref{fig:bonn_check}.  There is a trend of increasing $\dot\gamma_c$ up to a point where it reaches a plateau for $d\ge1.8$ mm.  The vast majority of rheometer measurements are done with $d< 1.8$ mm.  At larger gaps, samples tend to spill easily because they are comparable to the capillary length of water (2.7 mm) so the surface tension of the solvent is not enough to hold the sample in place against gravity. Therefore, for our analysis we only use data at $d \le 1.8$ mm.

To obtain a fit function for $d\le 1.8$ mm, we assume  a power law relationship
\begin{equation}
\dot\gamma_c \propto d^{1/B_d} \ .
\label{eqn:gapdependence}
\end{equation}

\noindent  Since we already obtained a fit coefficient $A$ in Eq.~\ref{eqn:gapdependence}, a proportionality coefficient is not needed here when combining the expressions to obtain $\phi_{eff}(\dot\gamma_c,d)$.  This eliminates the need to account for differences such as  packing fraction or solvent viscosity that would affect the value of that proportionality.  A least squares fit of Eq.~\ref{eqn:gapdependence} to the data for $d \le 1.8$ mm in Fig.~\ref{fig:bonn_check} yields $B_d = 0.284\pm0.022$ with the uncertainty in $\dot\gamma_c$ adjusted to 20\% to obtain a reduced $\chi^2=1$.  This value of $B_d$ is consistent with the values of $B=0.303\pm 0.038$ and $B=0.241\pm0.031$ obtained in the fits of Eq.~\ref{eqn:phidependence}, indicating that $\dot\gamma_c$ scales with both $d$ and $\phi_c-\phi_{wt}$ in the same way. 


\subsection{$\phi_{eff}(\dot\gamma_c,d)$}
\label{sec:combined_model}

An expression for $\phi_{eff}(\dot\gamma_c, d)$ can be obtained by combining the relations for $\phi_{eff}(\dot\gamma_c)$ and $\dot\gamma_c(d)$ from Eqs.~\ref{eqn:phidependence} and Eq.~\ref{eqn:gapdependence}.  Since the exponents $B$ and $B_d$ are consistent with each other, we set $B_d=B$ to  obtain a simpler expression:
\begin{equation}
\phi_{eff} = \phi_c- \frac{A'}{d}\dot\gamma_c^B \ .
\label{eqn:combined_model}
\end{equation}

\noindent To obtain values of $A'$ and $B$ for a general model, we simultaneously fit Eq.~\ref{eqn:combined_model} to the data for both gaps in Fig.~\ref{fig:phi_cricSR_all}.  Adjusting the input error on $\phi_{wt}$ to 0.010 to obtain a reduced $\chi^2=1$ yields $B=0.268\pm0.018$ and $A' =0.0290\pm 0.0020$ when $\dot\gamma_c$ is in units of s$^{-1}$ and $d$ is in units of mm.  Corresponding model curves at fixed $d$ that correspond to the data in Fig.~\ref{fig:phi_cricSR_all} are plotted as red lines in that figure.  These curves are seen to agree extremely well with the fits of Eq.~\ref{eqn:phidependence} without the dependence on gap $d$. The error here of 0.010 is the average of the errors on the fits of the individual curves in Fig.~\ref{fig:phi_cricSR_all}, so Eq.~\ref{eqn:combined_model} captures the $d$-dependence of that data accurately without any additional error.  This agreement confirms the validity of the assumption that the different materials used by Fall et al.~\cite{FBOB12} have the same scaling for $\phi_{eff}(d)$. 

$\phi_{eff}$ can now be calculated from Eq.~\ref{eqn:combined_model} for any suspension of cornstarch and water where $\dot\gamma_c$ is measured.  The usefulness of resolving data near $\phi_c$ can be seen in Fig.~\ref{fig:viscosity_shearrate_all}.  The viscosity curves plotted in Fig.~\ref{fig:viscosity_shearrate_all} are ordered by decreasing $\phi_{eff}$ from upper left to lower right.  In contrast, data taken using the same methods plotted in terms of $\phi_{wt}$ are not well-ordered, due to the uncertainty of 0.01 in $\phi_{wt}$ \cite{MB17}.

While it could be insightful to normalize $\dot\gamma_c$ in Eq.~\ref{eqn:combined_model} to rewrite $A'$ in terms of a length, to our knowledge, there is no timescale relevant to shear thickening that is independent of packing fraction that would allow us to do this.

\section{Improved precision and range of applicability}
\subsection{Random errors}
\label{sec:scatter}

\begin{figure}
\centering
\includegraphics[width=0.45\textwidth]{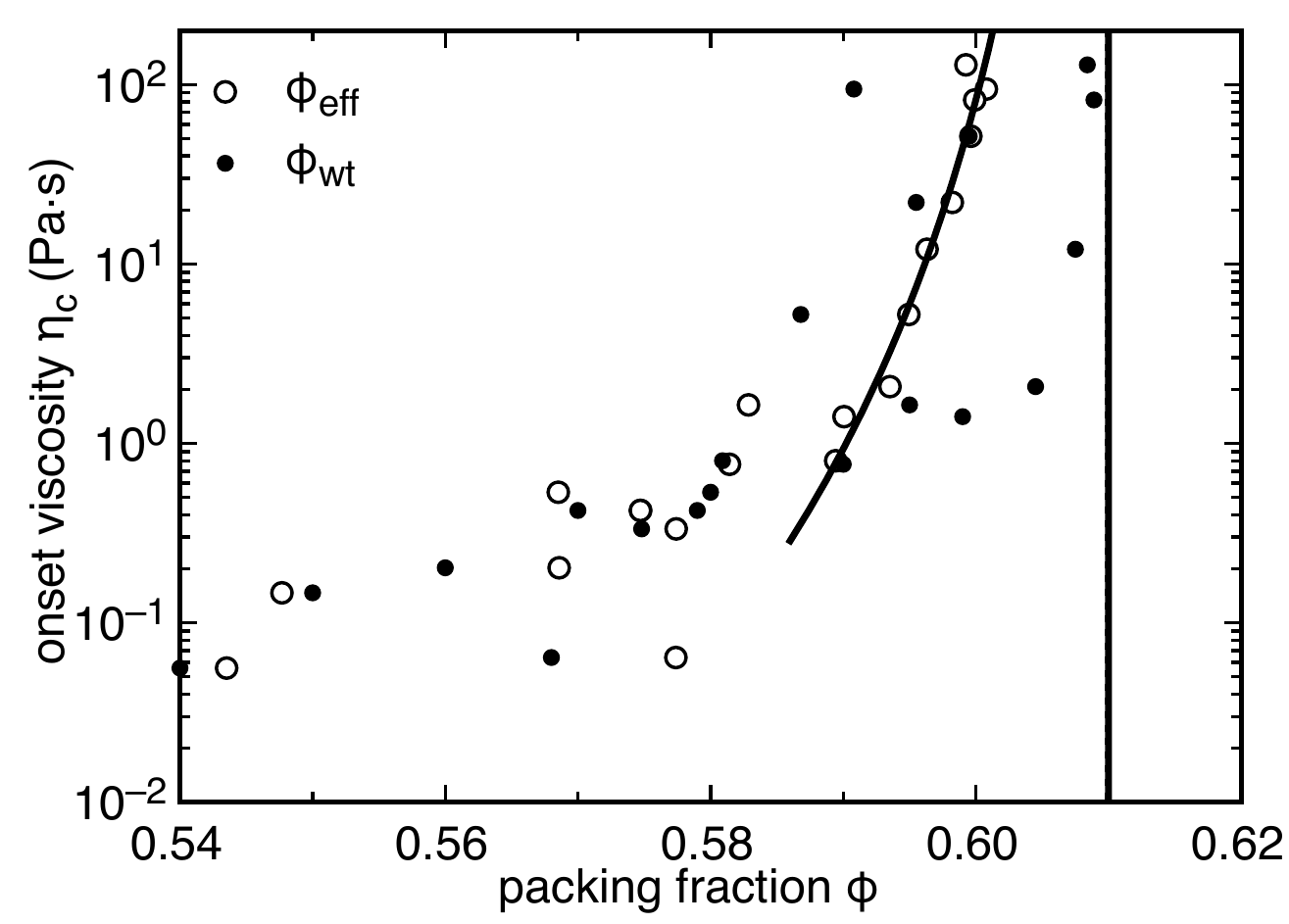}
\caption{Viscosity $\eta_c$ at the onset of DST as a function of packing fraction. Open symbols: directly measured weight fraction $\phi_{wt}$. Solid symbols: effective packing fraction $\phi_{eff}$ using Eq.~\ref{eqn:combined_model}. Dotted vertical line: critical packing fraction $\phi_c$. Solid line: fit of $\eta_c(\phi_{eff})$ for $\phi_{eff}\ge 0.586$.  The scatter in $\phi_{eff}$ is less than the scatter in $\phi_{wt}$ in this range.
}
\label{fig:viscosity_phi}
\end{figure}

Here we show how much the scatter is reduced by using effective packing fraction $\phi_{eff}$ compared to the directly measured weight fraction $\phi_{wt}$.    A  comparison can be made by plotting another quantity that varies strongly near $\phi_c$, as small errors in $\phi$ would be clearly apparent.  Specifically, we use the viscosity $\eta_c$ at the onset of DST  as defined in Sec.~\ref{sec:dotgammac}.  In Fig.~\ref{fig:viscosity_phi}, we plot the onset viscosity $\eta_c$  vs.~both $\phi_{wt}$ and $\phi_{eff}$ for $d=1.25$ mm.  It can be seen that there is less scatter for $\phi_{eff}$ than for $\phi_{wt}$  near the critical point.  To quantify the scatter, we fit a power law plus a constant $\phi_c=0.61$ to $\phi_{eff}(\eta_c)$ to the data, analogous to Eq.~\ref{eqn:phidependence} with $\eta_c$ in place of $\dot\gamma_c$.  This fit function has the expected divergence of $\eta_c(\phi_{eff})$ at $\phi_c$ \cite{KD59,BJ09}. We fit the inverted form to avoid problems with fit algorithms near a singularity.  We input a 14\% error on $\eta_c$ equal to the standard deviation of the mean, and adjust the error $\Delta\phi_{eff}$ on $\phi_{eff}$ to obtain a reduced $\chi^2=1$.  This fit is shown in Fig.~\ref{fig:viscosity_phi} for $\phi_{eff} > 0.586$.  For this range, we find the required input error $\Delta\phi_{eff}= 0.0008$.  This error corresponds to the root-mean-square (rms) difference from the fit, which includes sample-to-sample scatter plus any deviation of the fit function from the `true' function describing the data.  Thus, this uncertainty $\Delta\phi_{eff}$ reported is an upper bound on the random error of $\phi_{eff}$.  Alternatively, fitting $\eta_c(\phi_{wt})$ for the same data using the same method requires an error $\Delta\phi_{wt} =0.0082$.  The error is an order of magnitude smaller using $\phi_{eff}$.  

\begin{figure}
\centering
\includegraphics[width=0.45\textwidth]{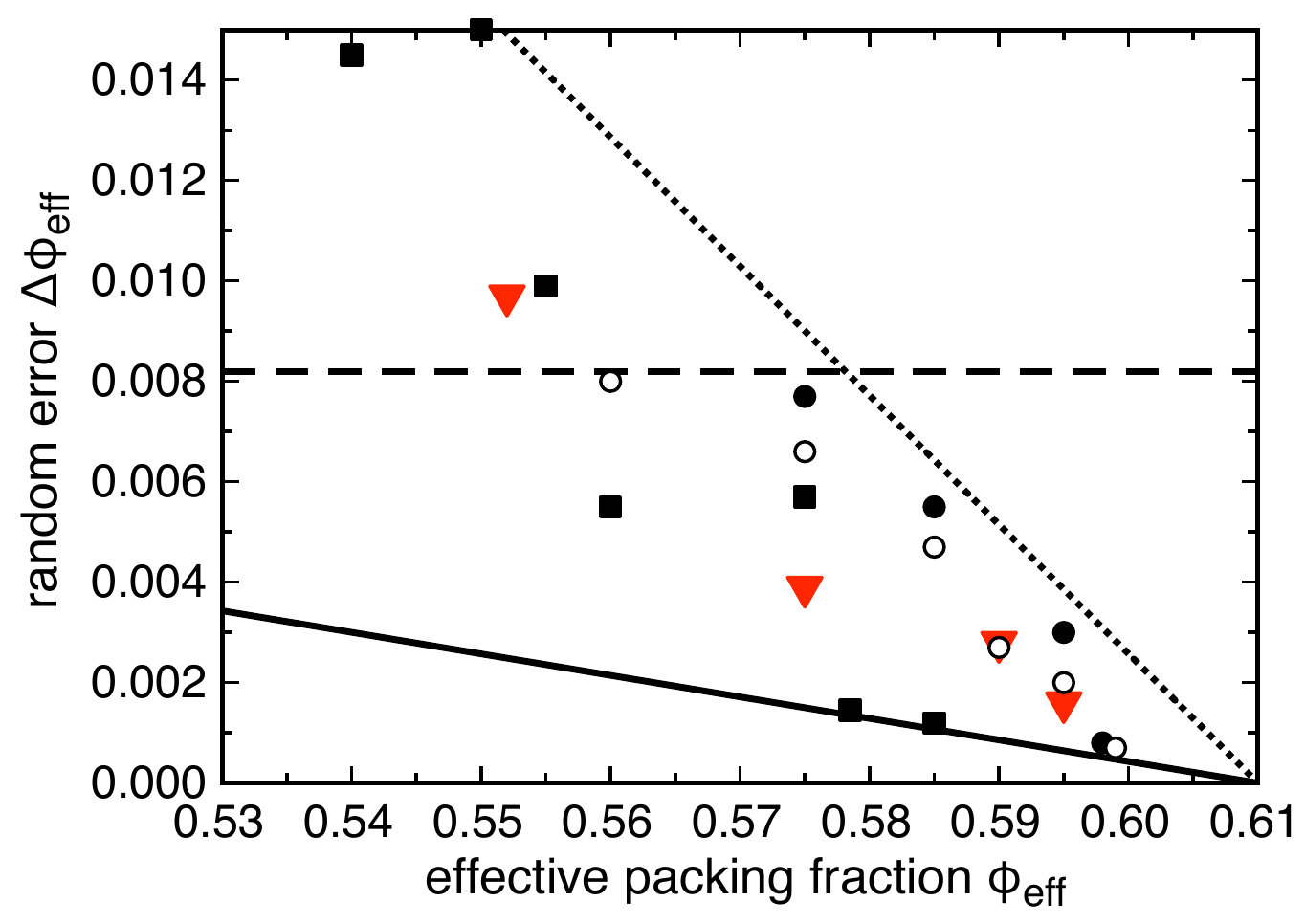}
\caption{(color online) Random error $\Delta\phi_{eff}$ as a function of $\phi_{eff}$ for different measurement and analysis methods.   Solid circles: $d=1.25$ mm.   Squares:  $d=0.61$ mm.  Open circles: $d=1.25$ mm, using a different analysis method to obtain $\dot\gamma_c$ \cite{MB17}.     Triangles:  data from a different laboratory with a different flow geometry, control mode, experimenter, and environmental conditions \cite{BJ09}.  Solid line: ideal random error on $\phi_{eff}$ from Eq.~\ref{eqn:propagated_error}. Dotted line:  6 times the ideal error as a guide to the eye. Dashed line:  error on $\phi_{wt}$.  Generally, the random error $\Delta\phi_{eff}$ is less than the error on $\phi_{wt}$ for $\phi_{eff} \stackrel{>}{_\sim} 0.56$, and decreases as $\phi_{eff}$ approaches $\phi_c$, with some variation in between 1-6 times the ideal propagated error.
}
\label{fig:scatter_phi_eff}
\end{figure}

   To illustrate how this uncertainty decreases as $\phi_{eff}$ approaches $\phi_c$ is, we plot the random error $\Delta\phi_{eff}$ for different fit ranges as a function of $\phi_{eff}$ in Fig.~\ref{fig:scatter_phi_eff}.   This is plotted as a function of the center of the fit range, where the upper end of the fit range is always fixed at $\phi_c$.   Results are shown for both $d=0.61$ mm (solid circles), $d=1.25$ mm (squares) in Fig.~\ref{fig:scatter_phi_eff}.  In each case, $\Delta\phi_{eff}$ tends to decrease as $\phi_{eff}$ approaches $\phi_c$, and is smaller than $\Delta\phi_{wt}$ for $\phi_{eff} \stackrel{>}{_\sim} 0.56$.   
   
Here we check if the random error $\Delta\phi_{eff}$ remains small when using different analysis methods.  In Ref.~\cite{MB17}, $\dot\gamma_c$ was obtained for each ramp individually, then these values were averaged over 4 ramps to obtain a value of $\dot\gamma_c$ for that $\phi_{wt}$.  Using this method, $\Delta\phi_{eff}$ is plotted is plotted as the open circles in Fig.~\ref{fig:scatter_phi_eff} for $d=0.61$  mm.  These results are in a similar range as solid circles which used the analysis technique explained in Sec.~\ref{sec:viscosity}.  This confirms the scatter in data is reduced significantly on the $\phi_{eff}$ scale compared to $\phi_{wt}$ even for different analysis methods. 


As the ultimate test of the ability of $\phi_{eff}$ to reduce scatter, we analyzed data under the most different experimental conditions we could obtain.   We use a dataset from a previous publication \cite{BJ09}. This dataset had a different experimenter, taken in a different laboratory, with different environmental conditions (relative humidity $40\pm 2\%$, air temperature $22.8\pm0.1^{\circ}$ C, and the rheometer controlled at $20.0^{\circ}$ C),  a different rheometer and flow geometry (cylindrical Couette with a gap of $d=1.13$ mm), a different measurement procedure (stress-controlled measurements with a ramp rate of 500 s/decade), and a different preshear (covering a wider range of stress above $\tau_{max}$ for all $\phi_{eff}$).  
 We did  the same fit of $\phi_{eff}(\eta_c)$, using a 10\% error on $\eta_c$ which represents the run-to-run standard deviation for that dataset \cite{BJ09}.  $\Delta\phi_{eff}$ is plotted as triangles in Fig.~\ref{fig:scatter_phi_eff} for different fit ranges.  $\Delta\phi_{eff}$ is similar to that found for the experiments described in Sec.~\ref{sec:methods}, and again decreases as $\phi_{eff}$ approaches $\phi_c$, and is smaller than the uncertainty on $\phi_{wt}$ for large $\phi_{eff}$ ($\Delta\phi_{wt}$ was 0.009 for this experimenter).  This confirms the scatter in the data is reduced significantly on the $\phi_{eff}$ scale compared to $\phi_{wt}$, regardless of experimenter, methods, or measurement conditions.  


 To illustrate why the error $\Delta\phi{eff}$ decreases near $\phi_c$, we calculate the ideal propagated error from Eq.~\ref{eqn:combined_model}
\begin{equation}
\Delta \phi_{eff} = (\phi_c-\phi_{eff})B\Delta\dot\gamma_c/\dot\gamma_c \ , 
\label{eqn:propagated_error}
\end{equation}

\noindent assuming that only the error $\Delta\dot\gamma_c$ on $\dot\gamma_c$ contributes to the random error on $\phi_{eff}$.  We plot this ideal $\Delta\phi_{eff}$ in Fig.~\ref{fig:scatter_phi_eff} for $B=0.268$ and  $\Delta\dot\gamma_c/\dot\gamma_c=0.16$, corresponding to the value for most of our data.  The measured random errors are somewhat larger than this prediction, in the range of 1-6 times the expected error in the DST range (the dotted line in Fig.~\ref{fig:scatter_phi_eff} shows 6 times the propagated error as a guide to the eye).  The difference between the measured and ideal propagated errors account for other sources of random error such as inherent irreproducibility of the sample, as well as the difference between the fit functions and true functions that describe $\eta_c(\phi_{eff})$ and $\dot\gamma_c(\phi_{eff})$, which is expected to increase farther from $\phi_c$ where the divergent scaling becomes less dominant. 

We also generally found the error $\Delta\phi_{eff}$ is smaller than the error on $\phi_{wt}$ for $\phi_{eff} \stackrel{>}{_\sim} 0.56$ in Fig.~\ref{fig:scatter_phi_eff}.   On the other hand, for $\phi_{eff} \stackrel{<}{_\sim} 0.55$, that means the error $\Delta\phi_{eff}$ is larger than the error on $\phi_{wt}$. Since the precision of $\phi_{eff}$ relies on the large slope of $\dot\gamma_c^{-1}$ near the critical point $\phi_c$, it is not surprising that it is less precise farther away from $\phi_c$.  Thus, we only recommend using $\phi_{eff}$ for $\phi_{eff} \stackrel{>}{_\sim} 0.56$, which corresponds to the DST range (shown in Fig.~\ref{fig:viscosity_shearrate_all}).

\subsection{Systematic errors}
\label{sec:systematicerror}

\begin{figure}
\centering
\includegraphics[width=0.45\textwidth]{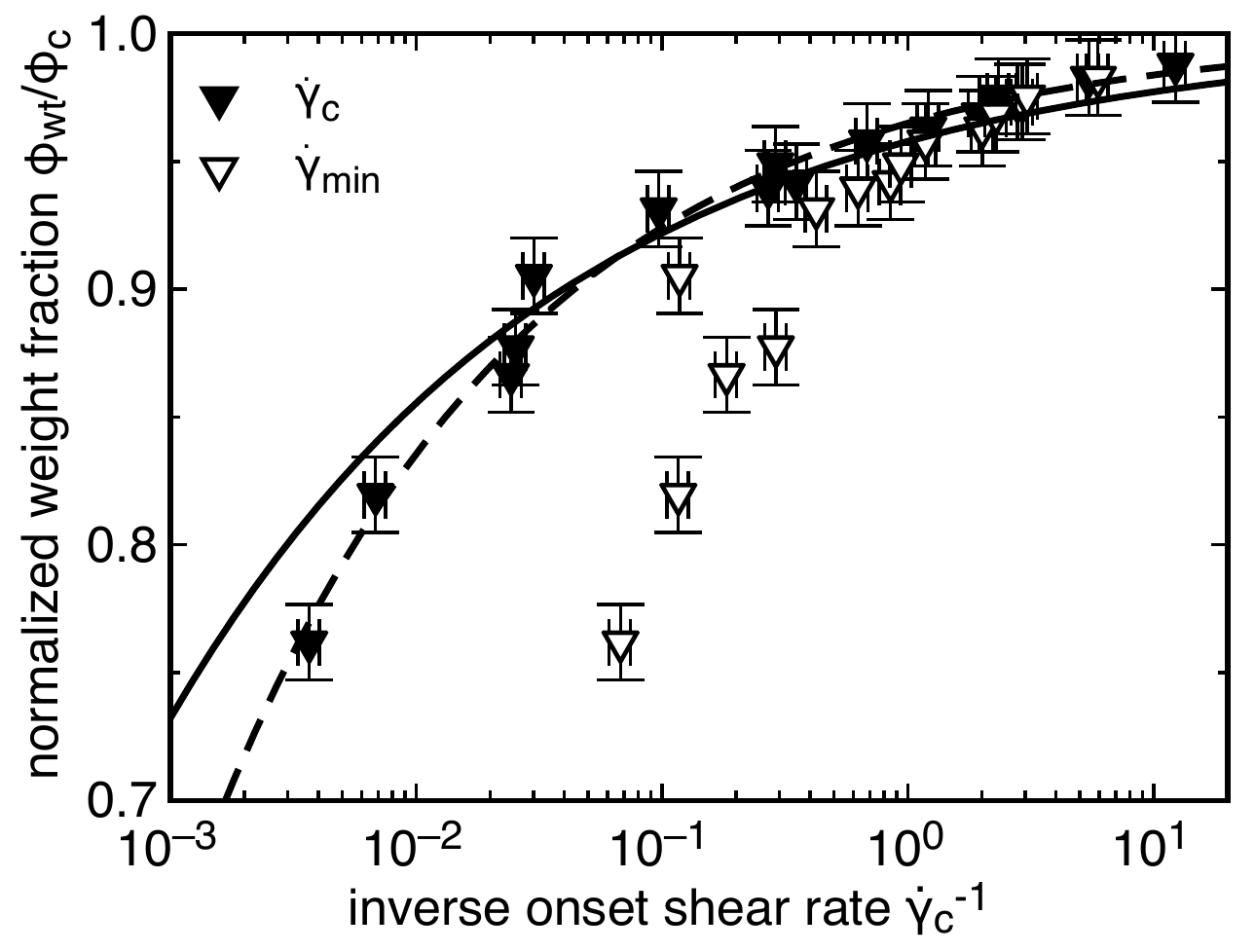}
\caption{Normalized weight fraction $\phi_{wt}/\phi_c$ as a function of inverse onset shear rate $\dot\gamma_c^{-1}$ for data taken in a different laboratory \cite{BJ09}.   Solid triangles: plotted as a function of $\dot\gamma_c^{-1}$ measured at the onset of DST.  Open triangles: plotted as a function of $\dot\gamma_{min}^{-1}$ measured at the onset of shear thickening.  Solid line: effective packing fraction $\phi_{eff}$ from Eq.~\ref{eqn:combined_model} with the parameters from Sec.~\ref{sec:combined_model}. Dashed line: best fit power law function to the solid triangles.  The model of Eq.~\ref{eqn:combined_model} differs from the best fit by a rms difference of 0.013 in $\phi_{eff}/\phi_c$, corresponding to a systematic error on $\phi_{eff}$ when comparing data from different labs with different temperature and humidity. The difference between the open and solid triangles indicates a much larger systematic error due to different definitions of the onset shear rate.
}
\label{fig:differentlabBJ09}
\end{figure}

Here we identify systematic error on $\phi_{eff}$, which is useful for comparing data from different labs, experimental procedures, equipment, environmental conditions, or experimenters.   
 As an example between very different datasets, we consider the dataset of Ref.~\cite{BJ09}, where the difference in $\phi_c$ based on $\phi_{wt}$ is 0.12.  Since $\phi_c$ is the critical point that controls the strength of DST, we hypothesize that this systematic error could be reduced if packing fractions are normalized by this critical point.  We plot this normalized $\phi_{wt}/\phi_c$ for the data from Ref.~\cite{BJ09} in Fig.~\ref{fig:differentlabBJ09} (solid triangles) as a function of $\dot\gamma_c^{-1}$.   For comparison, we plot Eq.~\ref{eqn:combined_model} with $d = 1.13$ mm as the solid line.  We exclude data from the quantitative analysis that were so close to the critical point that the yield stress shifted $\dot\gamma_c$ away from the diverging trend \cite{BJ09}.  The data of \cite{BJ09} agree closely with our $\phi_{eff}/\phi_c$, with a rms difference of 0.013 between the best fit of Eq.~\ref{eqn:phidependence} to the data in Fig.~\ref{fig:differentlabBJ09} and the parameters from Sec.~\ref{sec:combined_model}.  This corresponds to the systematic error due to all of the different measuring conditions and methods between the two experiments.  The small difference indicates that the same relationship $\phi_{wt}(\dot\gamma_c)/\phi_c$ or $\phi_{eff}(\dot\gamma_c)/\phi_c$ holds for experiments in different labs within a systematic error of 0.013.  This normalization is has also been found useful for collapsing shear thickening data even for particles of different materials or shapes \cite{BZFMBDJ11}.

The small systematic error reported above for different experiments assumes that  data are analyzed the same way.  If we calculate $\dot\gamma_c$ for each viscosity ramp before averaging (the method of Ref.~\cite{MB17}), we find a nearly identical systematic error on $\phi_{eff}/\phi_c$ of 0.014.   On the other hand, a larger systematic error could result from a different definition of the onset shear rate.  If we instead analyze the onset of shear thickening $\dot\gamma_{min}$ based on the lowest shear rate where $\eta(\dot\gamma)$ has a positive slope (as is often done), we obtain the open triangles in Fig.~\ref{fig:differentlabBJ09}.  There is a large systematic difference from $\phi_{eff}$ based on $\dot\gamma_c$ at the onset of DST.   Since other groups may record $\dot\gamma_{min}$ instead of $\dot\gamma_c$, we provide a conversion function to apply $\phi_{eff}$ for datasets in terms of $\dot\gamma_{min}$.  To obtain this conversion function,  we fit a power law to $\dot\gamma_c(\dot\gamma_{min})$ using the data of Ref.~\cite{BJ09}, again excluding data where $\dot\gamma_{c}$ is shifted by a yield stress.  This yields $\dot\gamma_c = (0.60\pm 0.04)\dot\gamma_{min}^{0.56\pm0.03}$.  This conversion can be applied before applying Eq.~\ref{eqn:combined_model}.  This extra conversion adds a systematic error of 17\% on $\dot\gamma_c$, and up to 0.004 on $\phi_{eff}/\phi_c$ in the DST range based on a propagation of the fit errors from Eq.~\ref{eqn:propagated_error}.  
  
Since the normalized scale $\phi_{eff}/\phi_c$ has smaller systematic errors when comparing data from different labs, it is desirable to present packing fraction data on this scale.  The normalized version of Eq.~\ref{eqn:combined_model} is
\begin{equation}
\frac{\phi_{eff}}{\phi_c} = 1- \frac{\tilde A'}{d}\dot\gamma_c^B \ ,
\label{eqn:phi_eff_rescaledl}
\end{equation}

\noindent where $\tilde A' = 0.0475$, and $B=0.268$  for $\dot\gamma_c$ is in units of s$^{-1}$ and $d$ is in units of mm.

\section{Summary} 

In this methods paper, we presented a technique to reduce uncertainty in measurements when there is a divergence in another quantity at a critical point, using the specific example of an effective packing fraction $\phi_{eff}$ of shear thickening suspensions of cornstarch and water.  The empirically fit conversion function is $\phi_{eff}/\phi_c = 1 -\tilde A'\dot\gamma_c^B/d$ where $B=0.268$, and $\tilde A' =0.0475$ when $\dot\gamma_c$ is in units of s$^{-1}$ and $d$ is in units of mm, and applies for $d\le 1.8$ mm (Figs.~\ref{fig:phi_cricSR_all}, \ref{fig:bonn_check}).  Obtaining $\phi_{eff}/\phi_c$ for a sample requires only a measurement of $\dot\gamma_c$ at the onset of DST and the rheometer gap size $d$, and plugging into the function $\phi_{eff}/\phi_c(\dot\gamma_c,d)$.   If the shear rate $\dot\gamma_{min}$ is measured at the onset of shear thickening instead of $\dot\gamma_c$ at the onset of DST, then $\dot\gamma_c$ can be obtained from $\dot\gamma_c = 0.60\dot\gamma_{min}^{0.56}$ before applying Eq.~\ref{eqn:combined_model}.

The main  advantage of the effective packing fraction is that the random error $\Delta\phi_{eff}$ is smaller than the error $\Delta\phi_{wt}$ on packing fractions measured by weight for $0.92 \stackrel{>}{_\sim}  \phi_{eff}/\phi_c < 1$ (Figs.~\ref{fig:viscosity_phi}, \ref{fig:scatter_phi_eff}), corresponding to the packing fraction range of DST (Fig.~\ref{fig:viscosity_shearrate_all}).  Because of the divergence of $\dot\gamma_c^{-1}$ at the critical point, the error on $\phi_{eff}$ gets even smaller closer to the critical point (Fig.~\ref{fig:scatter_phi_eff}).  This allows observations of trends over the narrow packing fraction range where DST occurs.  Our recent report of two distinct anomalous relaxation times in the range $0.97 < \phi_{eff}/\phi_c < 1$ is an example in which trends could not be clearly observed when plotting data as a function of $\phi_{wt}$ with errors of 0.013 in $\phi_{wt}/\phi_c$ (40\% of the measurement range), but trends could be resolved in terms of $\phi_{eff}$ \cite{MB17}.  

A second advantage of $\phi_{eff}/\phi_c$ is that it has a small systematic error of 0.013 when applied to datasets taken under different conditions, (i.e.~different labs, equipment, measurement techniques, experimenters, temperature and humidity) a significant improvement on the systematic errors on the order of $\approx 0.1$ in  $\phi_{wt}$ (Fig.~\ref{fig:differentlabBJ09}).  This small error on $\phi_{eff}/\phi_c$ makes it possible to compare data from different labs or seasons with high precision in the DST range.

Finally, while we used the specific example of cornstarch and water, this method to reduce uncertainties is expected to work in other suspensions with different fit parameters, as well as different critical phenomena where a measurable quantity diverges at a critical point.  To apply this method to other suspensions requires fitting Eq.~\ref{eqn:combined_model} to $\phi_{wt}(\dot\gamma_c,d)$ to obtain the fit parameters $A'$ and $B$.   While the value of $B$ is related to the exponent in the Krieger-Doherty relation, and we find consistent values for $B$ for suspensions in different solvents (e.g.~comparing to the data of Fall et al.~\cite{FBOB12}), the value of $B$ may be expected to depend on geometric properties of the particles like shape and roughness \cite{BJ09}.  For different critical phenomena, this approach relies on measuring a quantity that diverges at that critical point, and using that as a proxy to characterize the system.  A fit function can convert this to an effective parameter of choice for ease of interpretation.  The propagated errors are generally expected to be smaller on the effective scale because the error on a diverging quantity shrinks dramatically when propagated back to a non-diverging scale.

\section{Acknowledgments}

R. Maharjan and E. Brown both contributed significantly to the project design, collection of data, analysis, and writing of this manuscript.  Figures 1-4 were primarily the work of R. Maharjan, and Figs.~5-6 were  the work of E. Brown.  We thank Thomas Postiglione for the idea to test the analysis on data from a different lab.  This work was supported by the NSF through Grant No.~DMR 1410157.


\end{document}